\documentclass[10pt,british,tightenlines,eqsecnum,floats,aps,amsmath,amssymb,nofootinbib,superscriptaddress,prd,showpacs,showkeys]{revtex4}
\usepackage[utf8]{inputenc}
\usepackage{pdflscape}
\usepackage{babel}
\usepackage{graphicx}
\usepackage{esint}
\usepackage{inputenc}
\usepackage{amsfonts,amsmath,amssymb}

\usepackage[unicode=true,pdfusetitle,
 bookmarks=true,bookmarksnumbered=false,bookmarksopen=false,
breaklinks=false,pdfborder={0 0 1},backref=false,colorlinks=false]
 {hyperref}

\makeatletter
\@ifundefined{textcolor}{}
{%
 \definecolor{BLACK}{gray}{0}
 \definecolor{WHITE}{gray}{1}
 \definecolor{RED}{rgb}{1,0,0}
 \definecolor{GREEN}{rgb}{0,1,0}
 \definecolor{BLUE}{rgb}{0,0,1}
 \definecolor{CYAN}{cmyk}{1,0,0,0}
 \definecolor{MAGENTA}{cmyk}{0,1,0,0}
 \definecolor{YELLOW}{cmyk}{0,0,1,0}
}

\usepackage{epsfig}

\makeatother

\begin{document}

\title{\bf Generating CP violation from a modified Fridberg-Lee model}

\author{N. Razzaghi}
\email{n.razzaghi@qiau.ac.ir}

\affiliation{Department of Physics, Qazvin Branch, Islamic Azad
University, Qazvin, Iran}

\author{S. M. M. Rasouli}

\email{mrasouli@ubi.pt}

\affiliation{Departamento de F\'{i}sica,
Centro de Matem\'{a}tica e Aplica\c{c}\~{o}es (CMA-UBI),
Universidade da Beira Interior,
 Rua Marqu\^{e}s d'Avila
e Bolama, 6200-001 Covilh\~{a}, Portugal.}

\affiliation{Department of Physics, Qazvin Branch, Islamic Azad
University, Qazvin, Iran}

\author{P. Parada}

\email{pparada@ubi.pt}

\affiliation{Departamento de F\'{i}sica,
Centro de Matem\'{a}tica e Aplica\c{c}\~{o}es (CMA-UBI),
Universidade da Beira Interior,
 Rua Marqu\^{e}s d'Avila
e Bolama, 6200-001 Covilh\~{a}, Portugal.}

\author{P. V. Moniz}

\email{pmoniz@ubi.pt}

\affiliation{Departamento de F\'{i}sica,
Centro de Matem\'{a}tica e Aplica\c{c}\~{o}es (CMA-UBI),
Universidade da Beira Interior,
 Rua Marqu\^{e}s d'Avila
e Bolama, 6200-001 Covilh\~{a}, Portugal.}

\begin{abstract}
The overall characteristics of the solar and atmospheric neutrino
 oscillations are approximately consistent with a tribimaximal
form of the mixing matrix $U$ of the lepton sector.  Exact
tribimaximal mixing leads to $\theta_{13}=0$. However, the results
from the Daya Bay and RENO experiments have established, such that
in comparison to the other neutrino mixing angles, $\theta_{13}$
is small. Moreover, the atmospheric and solar mass splitting
differ by two orders of magnitude. These significant differences
constitutes the great enthusiasm and main motivation for our
research herein reported. Keeping the leading behavior of U as
tribimaximal. We would make a response to the following questions:
at some level, whether or not the small parameters such as the
solar neutrino mass splitting and $U_{e3}$, which vanish in a new
framework, can be interpreted as a modified FL neutrino mass
model? Subsequently, a minimal single perturbation leads to
nonzero values for both of them? Our minimal perturbation matrix
is constructed solely from computing the third mass eigenstate,
using the rules of perturbation theory. Let us point out that in
contrast with~\cite{my}, this matrix is not ad hoc assumed, but is
instead built following a series of steps we will outline. Also in
compared to the original FL neutrino mass model which generalize
it by inserting phase factors, our work is more accurate.
Subsequently, we produce the following results that add new
contributions to the literature: a) we obtain a realistic neutrino
mixing matrix with $\delta\neq0$ and $\theta_{23}=45^\circ$; b)
the solar mass splitting term is dominated by an imaginary term,
which could induce the existence of Majorana neutrinos, along with
explaining a large CP violation in nature; c) the ordering of the
neutrino masses is normal; however, at the end of the allowed
range, it becomes more degenerate ($97\%$); d) we also obtain the
allowed range of the mass parameters, which not only are in
accordance with the experimental data but also allow falsifiable
predictions for the masses of the neutrinos and the CP violating
phases which none of these results has been achieved in the
original FL neutrino mass model. Finally, let us emphasize that
the results retrieved through the framework proposed herein, as
compared to those produced in~\cite{my}, are much more efficient
concerning the currently available experimental data (namely, the
best fit column).
\end{abstract}


\date{2 July 2022}


\maketitle

\section{Introduction}
\label{int}

\label{int}

\indent One of the remarkable observational achievements
associated with neutrinos has been reported by neutrino
oscillation experiments~\cite{exp1, exp2, exp3, exp4,
information}, which establish the non-zero neutrino masses.
Concretely, that data yields information regarding neutrino masses
and mixing, which can be summarized as in
Table~\ref{h1}~\cite{information}.

\begin{table*}[t]
\begin{center}
\begin{tabular}{|c|c|c|}
\hline
Parameter &  The experimental data  $3\sigma$ range   & The best fit ($\pm1\sigma$)\\
\hline
$\Delta m_{21}^{2}[10^{-5}eV^{2}]$ & $6.94-8.14$ & $7.30-7.72$  \\
\hline
$|\Delta m_{31}^{2}|[10^{-3}eV^{2}]$& $2.47-2.63$ & $2.52-2.57$ \\
~&$2.37-2.53$&$2.42-2.47$ \\
 \hline
$\sin^{2}\theta_{12}$  & $3.02-3.34$ & $0.292-0.317$  \\
\hline
$\sin^{2}\theta_{23}$& $0.434-0.610$ & $0.560-0.588$  \\
~&$0.433-0.608$&$0.561-0.568$\\
\hline
$\sin^{2}\theta_{13}$& $0.02000-0.02405$ & $0.02138-0.02269$\\
~&0.02018-0.02424&$0.02155-0.02289$ \\
\hline
$\delta$& $128^\circ-359^\circ$ & $172^\circ-218^\circ$ \\
~&$200^\circ-353^\circ$&$256^\circ-310^\circ$\\
\hline
\end{tabular}
\caption{The experimental data for the neutrinos mixing
parameters. When multiple sets of allowed ranges are stated, the
upper row corresponds to normal hierarchy and the lower row to
inverted hierarchy.}\label{h1}
\end{center}
\end{table*}

In the standard parametrization, the lepton mixing matrix  is
given by \cite{mixing1, mixing2, mixing3},
\begin{equation}\label{emixing}
U_{PMNS}=\left(\begin{array}{ccc}c_{12}c_{13} & s_{12}c_{13} & s_{13}e^{-i\delta}\\
-s_{12}c_{23}-c_{12}s_{23}s_{13}e^{i\delta} &
c_{12}c_{23}-s_{12}s_{23}s_{13}e^{i\delta} &
s_{23}c_{13}\\s_{12}s_{23}-c_{12}c_{23}s_{13}e^{i\delta}
& -c_{12}s_{23}-s_{12}c_{23}s_{13}e^{i\delta} & c_{23}c_{13}\end{array}\right)\left(\begin{array}{ccc} e^{i\rho} & 0 & 0 \\
0 & 1 & 0\\0 & 0 & e^{i\sigma}\end{array}\right),
\end{equation}
where $c_{ij}\equiv\cos\theta_{ij}\text{ and
}s_{ij}\equiv\sin\theta_{ij}$ [for $i,j=(1,2), (1,3) \text{ and }
(2,3) $]. The phase $ \delta $ is called the Dirac phase,
analogous to the CKM phase, and the phases $\rho$ and $ \sigma$
are called the Majorana phases and are relevant for Majorana
neutrinos.

Experimental results have therefore shown that $\theta_{13}$ does
not vanish but is very small in comparison to the other neutrino
mixing angles. This recent observation ushered the possibility of
leptonic CP violation, although the CP violating phase $\delta$ is
not a well measured quantity. Furthermore, there is not any data
about the magnitude of the Majorana phases $\rho$ and $\sigma$.
Moreover, as it has been shown, the solar mass splitting is about
two orders smaller than the atmospheric one. The sign of the
atmospheric mass splitting has not been determined yet. Therefore,
the query is that whether the neutrino mass spectrum either does
respect the normal ordering or does obey the inverted ordering.
Moreover, the absolute neutrino mass scale is unknown.
Theoretically, an important question is that how we can define
this distinguished neutrino mixing pattern such that it would be
perfectly feasible to obtain probable values of unknown parameters
along with the other measured ones.

The tribimaximal neutrino mixing matrix $U_{TBM}$ is one of the
well-known neutrino mixing matrices \cite{TBM1, TBM2, TBM3}, which
is given by
\begin{equation}\label{etbm}
U_{TBM} =\left(\begin{array}{ccc}-\sqrt{\frac{2}{3}} & \frac{1}{\sqrt{3}} & 0\\
\frac{1}{\sqrt{6}} & \frac{1}{\sqrt{3}} &
-\frac{1}{\sqrt{2}}\\\frac{1}{\sqrt{6}} & \frac{1}{\sqrt{3}} &
\frac{1}{\sqrt{2}}\end{array}\right).
\end{equation}
The general exact tribimaximal mixing matrix $U_{TBM}$, regardless
of the model, fixes the element $(U_{e3})_{TBM}=0$. It is
important to note that, with the exception of $\theta_{13}$, the
values of the mixing angles associated with the matrix $U_{TBM}$
are consistent with the data of Table \ref{h1}. The role of a
non-zero $\theta_{13}$, or equivalently $U_{e3}$, is rather
relevant to many concepts in the lepton sector. It is necessary
for CP violation in neutrino oscillations and may be necessary to
explain leptogenesis. For CP violation, of course, both
$\theta_{13}$ and the complex phase $\delta$ should be non-zero.
Moreover, $\theta_{13}\neq0$ mirrors an equivalent feature present
in the quark sector, where mixing between all three generations is
a confirmed result, although the mixing angles in the two sectors
are very different. However, the discovery of the $\theta_{13}$,
whose smallness (in comparison to the other mixing angles)
proposes to modifythe the neutrino mixing matrix by means of a
small perturbation about the basic tribimaximal structure.
Consequently, it can lead to a realistic neutrino mixing matrix.
There are many neutrino mass models~\cite{81, 82, 83, 84, 85,855}
which can assist us to obtain the tribimaximal neutrino mixing
matrix. In order to produce $\theta_{13}\neq0$ starting from an
initial tribimaximal structure, different approaches have been
investigated: In \cite{HeZee, final}, a perturbative analysis has
been examined, from $U_{TBM}$. In~\cite{Grinstein}, an alternative
method has been proposed in which a sequential `integrating out'
of heavy neutrino states is involved. The authors of
\cite{unitary1, unitary2, unitary3} have employed the approach of
parametrizing the deviation from the tribimaximal form. In
\cite{Boudjemaa1,Boudjemaa2, Boudjemaa3, Boudjemaa33, Boudjemaa4},
the deviations from tribimaximal mixing due to charged lepton
effects and Renormalization Group running have been the direction
of study. Alternative explorations, based on the $A(4)$ symmetry,
have been carried out in \cite{A4pert1,A4pert2, A4pert3, A4pert4,
Adhikary}, while in \cite{Delta, Group1, Group2, Group3, Group4},
other discrete symmetries have been the basis for investigations.

The mixing parameters and the mass ordering in Table~\ref{h1} are
required inputs for recognizing viable models associated to the
neutrino masses. A natural option could be to take the mixing
angles as either $\theta_{13}=0$ or $\theta_{23}=\frac{\pi}{4}$,
such as those taken to obtain $U_{TBM}$ in Eq.~(\ref{etbm}), and
the solar mass splitting is missing. In this regard, one proposal
is provided as: which the atmospheric mass splitting and the
maximal mixing (of this sector) arise from a unperturbed mass
matrix while the smalle solar mass splitting and realistic
$U_{13}$ , ($\theta_{13}$ and $\delta$), are generating by a
perturbation. Moreover, it applies minor amendments to
$\theta_{12}$. By employing different methods in widely contexts,
a lot of endeavors have been pursued to generate some of the
neutrino parameters in perturbation theory \cite{P1}.

The purpose of our work is to introduce a framework, that
constitutes a modification of the neutrino mass setup proposed by
Friedberg and Lee (FL) \cite{FL}. The FL setting\footnote{A short
review of the Fridberg-Lee model is presented in
Appendix\ref{App.A}.} can be regarded as a successful
phenomenological neutrino mass model with flavor symmetry, which
can be appropriately and equivalently employed for both Dirac and
Majorana neutrinos\rlap.\footnote{To our knowledge, there is no
strong evidence regarding the identity of neutrinos, which could
be of the Majorana or the Dirac type.}

 However, in our work, we will
 employ instead a fundamental approach, which is different from that used in~\cite{my}.
Let us be more precise. In our herein paper, the perturbation mass
matrix will not be added by hand, but in contrast, it will be
thoroughly computed within a series of steps. To this aim, we will
be using the third perturbed mass eigenstate within perturbation
theory methods. More concretely, the perturbation mass matrix will
be thus constructed. We proceed as follows: (i) By employing
perturbation theory in the mass basis with real parameters, we
obtain the elements of the perturbation matrix which breaks the
$\mu-\tau$ symmetry. It will be seen that we get
$\theta_{13}\neq0$, but we do not have CP violation yet. (ii) We
extend our work to the case with CP violation, and show that a
complex perturbation matrix will be generated. In this case we
have nonzero values for both $\theta_{13}$ and $\delta$. We also
investigate the solar neutrino mass splitting in which an
imaginary term will be dominant and lead to the generation of the
Majorana phases. Consequently, we obtain CP violation along with a
realistic neutrino mixing matrix. (iii) Finally, by comparing our
phenomenological results with the corresponding experimental data,
we will set up allowed parameter ranges, along with neutrino
masses and CP violation phases.

This paper is organized as follows. In the next section we briefly
introduce our (modified FL) model, and then present the results of
the real and complex perturbation analysis described above in two
subsections, separately.

Moreover, for the complex case we compute the perturbation mass
matrix generating CP violation and we get a realistic neutrino
mixing matrix. In section \ref{exp}, we map two of the
experimental data onto the allowed region of our parameter space.
Thereafter, we find the presently allowed ranges for all the
parameters (especially perturbation parameters) of the model.
Finally, not only do we check the consistency of all of the
results with the available experimental data, but we also present
our predictions for the actual masses and CP violation parameters.
In section~\ref{concl} we summarize and analyze the results. In
Appendix~\ref{App.A}, we briefly introduce the FL model.

\section{Modified Friedberg-Lee model}
\label{Set up}
 In this section we construct our model, within
the FL framework, based on the basic tribimaximal neutrino mixing
matrix. We compute the minimal neutrino mass perturbation matrix,
along with a realistic neutrino mixing matrix. It is important to
note that the innovative distinguishing characteristic of the
preset work is that a neutrino mass perturbation will be erected
from the minimal principle of the perturbation theory. Namely, we
will not add a neutrino mass perturbation from any ad hoc
assumptions, whilst in \cite{my}, the neutrino mass perturbation
was added by hand by considering a few symmetries. It is important
to note that, according to the experimental data reported in
table~\ref{h1}, the results of our herein improved model indicate
an efficiency fitting increase of about $30$ percent in contrast
to those presented in~\cite{my}. In section~\ref{exp}, we will
further elaborate with more detail concerning this result.
 The tribimaximal neutrino mixing is a natural consequence of the
$M_{FL}$ mass matrix in the case of $\mu-\tau$ permutation
symmetry (namely, the neutrino mass matrix remains invariant under
interchanging indices $\mu$ and $\tau$ \cite{mutau}).
From Eq.~(\ref{efl}) it is apparent that $M_{FL} $ possesses exact
$\mu-\tau$ symmetry only when $b=c$. The magic property\footnote{
The sum of elements whether in every row or in every column of the
neutrino mass matrix is identical \cite{magic}.} of $M_{FL}$
obviously remains under exact $\mu-\tau$ symmetry. Setting $b=c$
and using the hermiticity of $M_{FL}$, a straightforward
diagonalization procedure yields $U^{T}M_{FL}U=\tilde{M}$, where
\begin{equation}\label{emm1}\vspace{.2cm}
\tilde{M} = \left(\begin{array}{ccc}3b+m_0 & 0 & 0\\
0 &  m_0 &0\\0 &0& 2a+b+ m_0
\end{array}\right)~~~~~~~~~~~~~~~~~\text{and}~~~~~~U=U_{TBM}.
\end{equation}
Also, we should notice that in the pure FL model one of the
neutrino masses is exactly zero. Moreover, just as in the general
FL setting $m_0$ must be positive ~\cite{FL}.

The reported experimental results have shown that the solar
neutrino mass difference is tiny and the inequality $\Delta
m^2_{21}>0$ is confirmed~\cite{information}. Considering the
experimental data, Eq.\,(\ref{emm1}) yields $b<0$, and $|b|\ll
m_0$. Therefore, we can set $\lambda=-b$ in Eq.\,(\ref{elanda})
and employ the transformations $a\rightarrow \alpha=a-b$ and
$b\rightarrow 0$. Consequently, in the flavor basis, the
unperturbed neutrino mass matrix is a magic $\mu-\tau$
symmetry\footnote{A magic $\mu-\tau$ symmetry for the mass matrix
is synonymous to a TBM mixing structure~\cite{FL, magic}. } and
given by
\begin{equation}\label{eunp}
M^{0}_{\nu} \simeq \left(\begin{array}{ccc}m_0 & 0 & 0\\
0 & \alpha + m_0 &-\alpha\\0 &-\alpha& \alpha + m_0
\end{array}\right),
\end{equation}
which has only two parameters $\alpha$ and $m_0$. Of course, the
diagonalized matrix which obtained from the mass matrix
$M^{0}_{\nu}$, Eq.\,(\ref{eunp}), is a special case of the mass
matrix Eq.\,(\ref{emm1}), which has three parameters $a$, $b$ and
$m_0$. It is of interest that they shares the same neutrino mixing
matrix given by Eq.\,(\ref{etbm})\footnote{$M^{0}_{\nu}$ mass
matrix in Eq.\,(\ref{eunp}) with $\mu-\tau$ symmetry and magic
symmetry can be diagonalized by $U_{TBM}$. Because of $U_{TBM}$ is
a effect of $\mu-\tau$ and magic symmetries in the neutrino mass
matrix. Therefore $U_{TBM}$ is in itself a mixture of those
symmetries. \cite{TTBM}}, $U_{TBM}~$\cite{FL}.

 The mass spectrum of $M^{(0)}_{\nu}$ is
 \footnote{Naturally, if $\alpha\ll m_0$, the neutrino masses would approach the
 quasi-degenerate regime.}
\begin{eqnarray}\label{emm2}\vspace{.2cm}
m^{(0)}_1 = m^{(0)}_2 =m_0,
~~~~\text{and}~~~m^{(0)}_{3}=2\alpha+m_0.
\end{eqnarray}
Here $m^{(0)}_1$ and $m^{(0)}_2$ are real and positive numbers,
but at this stage, the sign of $m^{(0)}_3$ is unknown. In the next
section, by comparing the results of our model with the
experimental data, we will see that
 the sign of $m^{(0)}_3$ as well as the ordering of it (with respect
 to $m^{(0)}_1$ and $m^{(0)}_2$) will be specified. We should mention
that, up to now, the shortcomings are: (i) the absence of the
solar neutrino mass splitting, (ii) the ordering of neutrino
masses is unknown and (iii) the mixing matrix is still $U_{TBM}$.
Thus, the main objective will be to obtain the solar mass
splitting by means of a mass perturbation, which is the cause of
$\theta_{13}\neq 0 $ and CP violation. Moreover, CP violation
conditions necessarily mandate that $\mu-\tau$ symmetry should be
broken. An interesting question is: after the $\mu-\tau$ symmetry
breaking, will $\theta_{23}=45^{\circ}$ remain valid or not?

In summary, up to now, we have proposed that the modified FL
neutrino mass matrix in Eq.\,(\ref{eunp}) has a combination in
which $\Delta m^2_{12}$ and  $\theta_{13}$ are vanishing, while
$\theta_{23}=\frac{\pi}{4}$. Moreover, the atmospheric mass
splitting $\Delta m^2_{31}$ does not vanish. Furthermore, the
solar mixing angle $\theta_{12}$ can be selected as chosen by the
popular mixing matrix as $U_{TBM}$. This is a good estimate of the
observed data although small characteristic are missing here.
Therefore, the neutrino mass matrix in Eq.\,(\ref{eunp}) has two
mass eigenvalues, $m^{(0)}_1$ and $m^{(0)}_2$, which
 are degenerate, hence it is highly distinctive from the original FL model, in which all neutrino
masses are different \cite{FL}. Moreover, to the best of our
knowledge, such kind of FL modification has not been performed in
the literature yet.

In the next stage, we will consider the attendance of a small
contribution,
 which can be obtained by employing the perturbation theory, which generates small parameters
 in the neutrino mixing component, namely, $U_{13}$, ($\theta_{13}$ and $\delta$), $\Delta m^2_{21}$
  and provides minor amendments to $\theta_{12}$ (but not to $\theta_{23}$).
  CP violation will be investigated. As previously
  mentioned, because of small $\theta_{13}$ and $\Delta m^2_{21}$,
we believe that the perturbative treatment is a more precise
method than others for getting the correction of the $U_{TBM}$. In
our point of view, our work is special even in the perturbative
treatment, because our minimal perturbation matrix is constructed
solely from computing the third mass eigenstate, using the rules
of perturbation theory, see Subsection A and B. Therefore, this
perturbation matrix could induce both $U_{13}$ and $\Delta
m^2_{21}$.
  It is worth mentioning, in the original FL model, by inserting phase factors in the neutrino mass matrix, the CP
  violation incorporate \cite{FL}.

In order to establish the structure of the neutrino mass matrix,
as noted before, our strategy is to employ the perturbation
theory. Thus, we set $M_\nu=M_\nu^{0}+M'_\nu$
 where $M'_\nu<<M_\nu^{0}$. In general, $M_\nu^{0}$ and $M'_\nu$
are symmetric and complex. However, as seen from
Eq.\,(\ref{eunp}), in this case $M_\nu^{0}$ is symmetric and real,
i.e. it is Hermitian. In the following two subsections we will
consider first the case where $M'_\nu$ is real and then the case
where it is complex, respectively. In either situation we have
$\theta_{13}\neq0$, but CP is conserved when $M'_\nu$ is real.
Furthermore, in the complex case the solar neutrino masses are
split and CP is violated.

 In the mass basis the
eigenstates of $M_\nu^0$ (the unperturbed mass eigenstates) are as
follows:

\begin{equation} \vspace{.2cm}\label{em0}
|\nu^{(0)}_{1}\rangle=\left(\begin{array}{ccc}1\\
0\\0
\end{array}\right), ~~~~|\nu^{(0)}_{2}\rangle=\left(\begin{array}{ccc}0\\
1\\0\end{array}\right), ~~~~~|\nu^{(0)}_{3}\rangle=\left(\begin{array}{ccc}0 \\
0\\1\end{array}\right),
\end{equation}
in which the first two mass eigenstates are degenerate. We choose
$M'_\nu$ such that $\nu_{1}$ and $\nu_{2}$ are its nondegenerate
eigenstates, namely,
$\langle\nu_{i}^{(0)}|M'_{\nu}|\nu_{j}^{(0)}\rangle=
m_{i}^{(1)}\delta_{ij}$ where $ (i,j=1,2)$, with $m_1^{(1)}\neq
m_2^{(1)}$. Then we take $(M'_\nu)_{33}=0$ and consequently need
to consider only $(M'_\nu)_{13}$ and $(M'_\nu)_{23}$. Therefore,
in order to reproduce the correct solar mixing, the basis vectors
$\nu_{1}$ and $\nu_{2}$ are chosen, while the physical basis is
fixed by the perturbation. It is straightforward to show that by
expressing the mass eigenstates given by Eq.\,(\ref{em0}) in terms
of the flavor basis, we can get the columns of $U_{TBM}$ as given
by  Eq.\,(\ref{etbm}). Consequently, in the flavor basis, the
eigenstates are given by

\begin{equation} \vspace{.2cm}\label{em00}
|\nu^{(0)}_{1}\rangle=\left(\begin{array}{ccc}-\sqrt{\frac{2}{3}}\\
\frac{1}{\sqrt{6}}\\\frac{1}{\sqrt{6}}
\end{array}\right), ~~~~|\nu^{(0)}_{2}\rangle=\left(\begin{array}{ccc}\frac{1}{\sqrt{3}}\\
\frac{1}{\sqrt{3}}\\\frac{1}{\sqrt{3}}\end{array}\right), ~~~~~|\nu^{(0)}_{3}\rangle=\left(\begin{array}{ccc}0 \\
-\frac{1}{\sqrt{2}}\\\frac{1}{\sqrt{2}}\end{array}\right).
\end{equation}
\subsection{CP conservation }
In this subsection, our aim is to determine the third perturbed
mass eigenstate in the CP conserving case. When this eigenstate is
expressed in the flavor basis, we must obtain the third column of
the neutrino mixing matrix, given by Eq.~(\ref{emixing}). Thus we
can compute the elements of the perturbation matrix
 which will also be employed in the next subsection.
 Again, it is necessary to mention that in this
method, we do not pick up a perturbation mass matrix by hand;
instead, we compute it using of the third perturbed mass
eigenstate, which is an unique feature and distinguishable from
that used in \cite{my}. As stated previously, here we assume
$M'_\nu$, which is symmetric, to also be real, and therefore
Hermitian. Hence, while it may generate a nonzero $\theta_{13}$,
it necessarily yields $\delta=0$, and so leads to no CP violation.
For the perturbation expansion we keep terms up to linear order in
$s_{13}$. To first order we have
\begin{equation}
|\nu_3\rangle=|\nu^{(0)}_3\rangle+\sum_{j \neq 3} {\cal
C}_{3j}|\nu^{(0)}_j\rangle \;, \label{eper1}
\end{equation}
where,
\begin{equation}
{\cal C}_{3j}=-{\cal
C}_{j3}=(m^{(0)}_3-m^{(0)}_j)^{-1}{<\nu^{(0)}_j|M_\nu^\prime|\nu^{(0)}_3>
} , \;\;\; (j \neq 3) . \label{eper2}
\end{equation}
In this case, the coefficients ${\cal C}_{3j}$ are real and
proportional to the elements ${(M'_\nu)}_{3j}$ in the mass basis.

Obviously, $|\nu_3\rangle$ in Eq.\,(\ref{eper1}) Should be equal
to the third column of the mixing matrix $U_{PMNS}$ (with
$\delta=0$) of Eq.\,(\ref{emixing}). In the flavor basis, by using
Eq.\,(\ref{eper1}),
 we can easily determine ${\cal C}_{31}$ and ${\cal C}_{32}$.
We obtain the matrix equation
\begin{equation}\label{e1}
\left(\begin{array}{ccc}s_{13}\\
s_{23}c_{13}\\c_{23}c_{13}
\end{array}\right)=\left(\begin{array}{ccc}\frac{-\sqrt{2}{\cal C}_{31}+{\cal C}_{32}}{\sqrt{3}}\\
-\frac{1}{\sqrt{2}}+\frac{{\cal C}_{31}}{\sqrt{6}}+\frac{{\cal
C}_{32}}{\sqrt{3}}\\\frac{1}{\sqrt{2}}+\frac{{\cal
C}_{31}}{\sqrt{6}}+\frac{{\cal
C}_{32}}{\sqrt{3}}\end{array}\right)
\end{equation}
To linear order in $s_{13}$, we obtain ${\cal
C}_{31}=-\sqrt{\frac{2}{3}} s_{13}$ and ${\cal
C}_{32}=\sqrt{\frac{1}{3}} s_{13}$, where we have used maximality
of the $2-3$ mixing angle, ($\theta_{23}=45^\circ$). Therefore, in
the mass basis, by using Eq.\,(\ref{emm2}) and Eq.\,(\ref{eper2}),
we have ${(M'_\nu)}_{13}=-2\alpha\sqrt{\frac{2}{3}} s_{13}$ and
${(M'_\nu)}_{23}=2\alpha\sqrt{\frac{1}{3}} s_{13}$.

Briefly, in the CP conserving case, we calculate solely
\begin{equation}\label{e111}
U=U_{TBM}+\left(\begin{array}{ccc}0 & 0 & s_{13}\\
\frac{-s_{13}}{\sqrt{3}} & \frac{s_{13}}{\sqrt{6}}
&0\\\frac{s_{13}}{\sqrt{3}} &\frac{-s_{13}}{\sqrt{6}}& 0
\end{array}\right)\footnote{U is unitary up to order $s_{13}$.}~~~~~~~~~\text{and}~~~~~~~~~M_\nu=M^{0}_\nu+\left(\begin{array}{ccc}0 & 0 & -2\alpha\sqrt{\frac{2}{3}} s_{13}\\
0 & 0 &2\alpha\sqrt{\frac{1}{3}}
s_{13}\\-2\alpha\sqrt{\frac{2}{3}} s_{13}
&2\alpha\sqrt{\frac{1}{3}} s_{13}& 0
\end{array}\right)
\end{equation}

\subsection{CP violation }
In this subsection, let us proceed our discussion, now addressing
CP violation. We now assume $M'_\nu$ to be a complex symmetric
matrix, thus not Hermitian; then this is also true for the total
mass matrix $M_\nu=M^0_\nu+M'_\nu$. This is accomplished by
considering nonzero values for both $\sin\theta_{13}$ and
$\delta$. The columns of the mixing matrix $U$ in
Eq.\,(\ref{emixing}) are eigenvectors of $M_\nu^\dagger M_\nu =
M_\nu^{0\dagger}M_\nu^0 + M_\nu^{0\dagger}M_\nu' +
M_\nu'^{\dagger}M_\nu^0$, where we have dropped the term ${\cal O}
(M_\nu')^2$. We should mention that the unperturbed term
$M_{\nu}^{0^{\dagger}}M_{\nu}^{0}$ is Hermitian, its eigenstates
are the columns of $U_{TBM}$ [in Eq.\,(\ref{etbm})], and its
eigenvalues are $|m_{1}^{(0)}|^{2}$, $|m_{2}^{(0)}|^{2}$, and
$|m_{3}^{(0)}|^{2}$. Instead of Eq.\,(\ref{eper2}), we now have
\begin{equation}
{\cal C}_{3j}= -{\cal C}^*_{j3}=\left(|m^{(0)}_3|^2 -
|m^{(0)}_j|^2\right)^{-1}{\cal M}_{j3}, \;\; (j \neq 3) \;
 \label{eper3}
\end{equation}
where ${\cal M}_{j3}={<\nu^{(0)}_j| ( M_\nu^{0\dagger}M_\nu' +
M_\nu'^{\dagger}M_\nu^0) |\nu^{(0)}_3> }$ and $|\nu_3\rangle$ is
reproduced, to first order, by substituting the expressions
associated to ${\cal C}_{3j}$ from Eq.\,(\ref{eper3}) into
Eq.\,(\ref{eper1}). Consequently, by using an appropriate variant
of Eq.\,(\ref{e1}) for this case, we get ${\cal C}_{31}=-\sqrt{2
\over 3}s_{13}e^{-i \delta}$ and ${\cal C}_{32} = \sqrt{1 \over
3}s_{13}e^{-i \delta}$. It is important to note that, in the mass
basis, due to the symmetric nature of $M'_\nu$ , it is easy to
relate ${\cal C}_{31}$ and ${\cal C}_{32}$ to the elements of
$M'_\nu$ as
\begin{eqnarray}
{\cal C}_{3j} \left(|m^{(0)}_3|^2 - |m^{(0)}_j|^2 \right)
&=&<\nu^{(0)}_j| ( M_\nu^{0\dagger}M_\nu' +
M_\nu'^{\dagger}M_\nu^0) |\nu^{(0)}_i>\nonumber\\&=& m^{(0)}_j
(M_\nu^\prime)_{j3} + m^{(0)}_3 (M_\nu^\prime)^*_{j3} ,\;\;   (j
\neq 3) \; . \label{eo3j}
\end{eqnarray}
Employing Eq.\,(\ref{eo3j}), we get $(M_\nu^\prime)_{13} =
-\sqrt{2 \over 3}\frac{|\Delta
m_{31}^{2}|~s_{13}}{g(\eta)}~e^{i\eta}$ and $(M_\nu^\prime)_{23}
=\sqrt{1 \over 3}\frac{|\Delta m_{31}^{2}|~s_{13}}{g(\eta)}~e^{i
\eta}$, where

$\Delta m_{31}^{2}=(m^{(0)}_3)^2-(m^{(0)}_1)^2$ is the atmospheric
mass splitting, (considering the expressions for
 $m^{(0)}_1$, $m^{(0)}_3$ from
Eq.\,(\ref{emm2})),
\begin{equation}
\eta = \tan^{-1}\left( \frac {\alpha+m_0}{\alpha}
\tan\delta\right),
 \label{edelta}
\end{equation}
and
\begin{equation}
g(\eta) = \left[m_0^2 + (2\alpha+m_0)^2 + 2m_0(2\alpha+m_0) \cos
2\eta \right]^{1/2} \;\;. \label{ephi}
\end{equation}
and the allowed range for both of $\eta$ and $\delta$ is $\{0,
2\pi\}$. From Eq.\,(\ref{ephi}), it can be seen that $|m^{(0)}_3|
-|m^{(0)}_1|\leq g(\eta)\leq|m^{(0)}_3| +|m^{(0)}_1|$.

Up to now, by using Eq.\,(\ref{eo3j}) and ${\cal C}_{31}$, we have
focused on deriving $\theta_{13} \neq 0$ via a perturbation
analysis starting from an FL setting and the basic tribimaximal
neutrino mixing matrix. Now, we investigate the solar neutrino
mass splitting. In our framework of minimal perturbation we take
$(M_\nu^\prime)_{12} = (M_\nu^\prime)_{21} = 0$. The first order
corrections to the neutrino masses are obtained from
$m^{(1)}_i\delta_{ij}= ~<\nu^{(0)}_i|M_{\nu}^\prime|\nu^{(0)}_j>$.
We consider these first-order mass corrections as
\begin{equation}
m^{(1)}_1 = m^{(1)}_3 = 0 ~~~{\rm and} ~~~m^{(1)}_2 \neq 0.
\label{efirstm}
\end{equation}

Therefore, in the mass basis, (\ref{efirstm}) implies that only
$(M_\nu^\prime)_{22} \neq 0$ whilst other diagonal elements of the
perturbation matrix vanish. Such a correction displays a nonzero
solar neutrino mass splitting in which $ m^{(1)}_2=m_2 - m_1 $,
and
 $\Delta m^2_{21} = (m_2)^2 -(m_1)^2$ takes positive values.
Consequently, in the mass basis, we obtain the final perturbation
matrix as
\begin{equation}\label{eM2}
M_\nu^\prime =~\left(\begin{array}{ccc}0 & 0 & -\sqrt{\frac{2}{3}}s_{13}F\\
0 & m_2^{(1)} &
\sqrt{\frac{1}{3}}s_{13}F\\-\sqrt{\frac{2}{3}}s_{13}F &
\sqrt{\frac{1}{3}}s_{13}F &0\end{array}\right),
\end{equation}
and
\begin{equation} F~=~\frac{|\Delta
m_{31}^{2}|}{g(\eta)} e^{i\eta}.~ \label{xdef}
\end{equation}

Now from the elements of $M_\nu^\prime $ in Eq. (\ref{eM2}), let
us define a dimensionless parameter as
$\varepsilon\equiv\frac{m_2^{(1)}g(\eta)}{|\Delta
m_{31}^{2}|s_{13}}$, which relates the solar mass splitting,
$m_2^{(1)}$, to $\sin\theta_{13}$. In the next section, we will
employ this parameter to obtain the order of $\sin\theta_{13}$. In
general, the solar mass splitting can take complex values, so let
us mention that the Majorana mass is given by $m^{(1)}_2 \equiv
|m^{(1)}_2| ~\exp(i \varphi)$. If we write $m_2 = m^{(0)}_2 +
m^{(1)}_2 \equiv |m_2| ~\exp(i \phi)$,
we obtain
\begin{equation}
|m_2| = \left[(m^{(0)}_1)^2 + (|m^{(1)}_2|)^2 + 2 m^{(0)}_1
|m^{(1)}_2| \cos\varphi\right]^{1/2}, \;\; \\ \phi =
\tan^{-1}\left[{|m^{(1)}_2|  ~\sin\varphi  \over m^{(0)}_1 +
|m^{(1)}_2|  ~\cos \varphi}\right]. \label{em16}
\end{equation}
Therefore, in the Majorana case, $\phi$ is the origin of the
Majorana phases which arise from the perturbation.
In the next section, we obtain interesting results associated to
$\varphi$ and $\phi$, namely, that $m_2^{(1)}$ is dominated by its
imaginary part, and so $\phi$ can take large values.

In order to relate our perturbation mass matrix to the FL model,
let us rewrite $M_\nu^\prime$, [which is given by Eq. (\ref{eM2})
and it was calculated in the mass basis, see Eq. (\ref{em0})] in
the flavor basis. Therefore, employing relation between mass and
flavor basis and rewrite $M_\nu^\prime$, as
\begin{equation}\label{eM3} M_\nu^{\prime(f)} =\frac{F
~s_{13}}{\sqrt{2}}\left(\begin{array}{ccc}0 &-1& 1\\
-1& 0 & 0\\1 &
0 &0\end{array}\right)+\frac{m_2^{(1)}}{3}\left(\begin{array}{ccc}1 & 1 & 1\\
1 & 1 & 1\\1 & 1 &1\end{array}\right).
\end{equation}
We observe that the first and the second terms on the right-hand
side are responsible for $\theta_{13}$ and for $\Delta m^2_{21}$,
respectively. Notice that $M_\nu^{\prime(f)}$ in Eq. (\ref{eM3})
violates both $\mu-\tau$ symmetry and the magic feature of the
total mass matrix. Using degenerate perturbation theory
\cite{Schiff}, to linear order in $s_{13}$, and the relation
associated to ${\cal C}_{3j}$ given by Eq. (\ref{eper3}), and
aware of ${\cal C}_{ij}=0$ if $i\neq3$, which we know from the
elements of $M_\nu^{\prime(f)}$ according to Eq. (\ref{eM2}), we
obtain the neutrino mixing matrix, with $\delta \neq 0$, as
\begin{equation}\label{eU}
U =U_{TBM}+\left(\begin{array}{ccc}0 & 0 & s_{13}e^{-i\delta}\\
-\sqrt{\frac{1}{3}}s_{13}e^{i\delta} &
\sqrt{\frac{1}{6}}s_{13}e^{i\delta} &
0\\\sqrt{\frac{1}{3}}s_{13}e^{i\delta} &
-\sqrt{\frac{1}{6}}s_{13}e^{i\delta} &0\end{array}\right).
\end{equation}
The nonzero $\delta$ indicates CP violation in the lepton sector.

In \cite{18}, with a different motivation in view, this same
 form for $U$ has been
discussed and the consistency with the observed mixing angles
noted.

A rephasing-invariant measure of CP violation in neutrino
oscillation is the universal parameter $J$ \cite{J} [given in
Eq.\,(\ref{eJ})], where its form is independent of the choice of
the Dirac or Majorana neutrinos.

Using Eqs.\,(\ref{edelta}) and\,(\ref{eU}), the expression for $J$
can be simplified as
\begin{eqnarray}\label{eJJ}\vspace{.5cm}
  J&=&-{1 \over 3
\sqrt{2}}s_{13} \sin \delta \nonumber\\&=&
 -{1 \over 3 \sqrt{2}}{s}_{13}
\frac{\left({\alpha \over \alpha+m_0}\right) \sin
\eta}{\sqrt{\cos^2\eta + \left({\alpha \over \alpha+m_0}\right)^2
\sin^2 \eta} }.
\end{eqnarray}
We should notice that in order to have CP violation in the lepton
sector, both $s_{13}$ and $\eta$ must take nonzero values.

\section{Comparison with experimental data}\label{exp}
In this section, we compare the results obtained
through our (modified FL) model with the experimental data
\cite{information, information1}. It is important to note that in
the original FL model \cite{FL} almost there is no numerical
prediction for neutrino parameters. Therefore, in this section, we
compare our herein results with the corresponding ones obtained in
\cite{my}. Let us perform this in two steps.

\begin{figure}[th] \includegraphics[width=13cm]{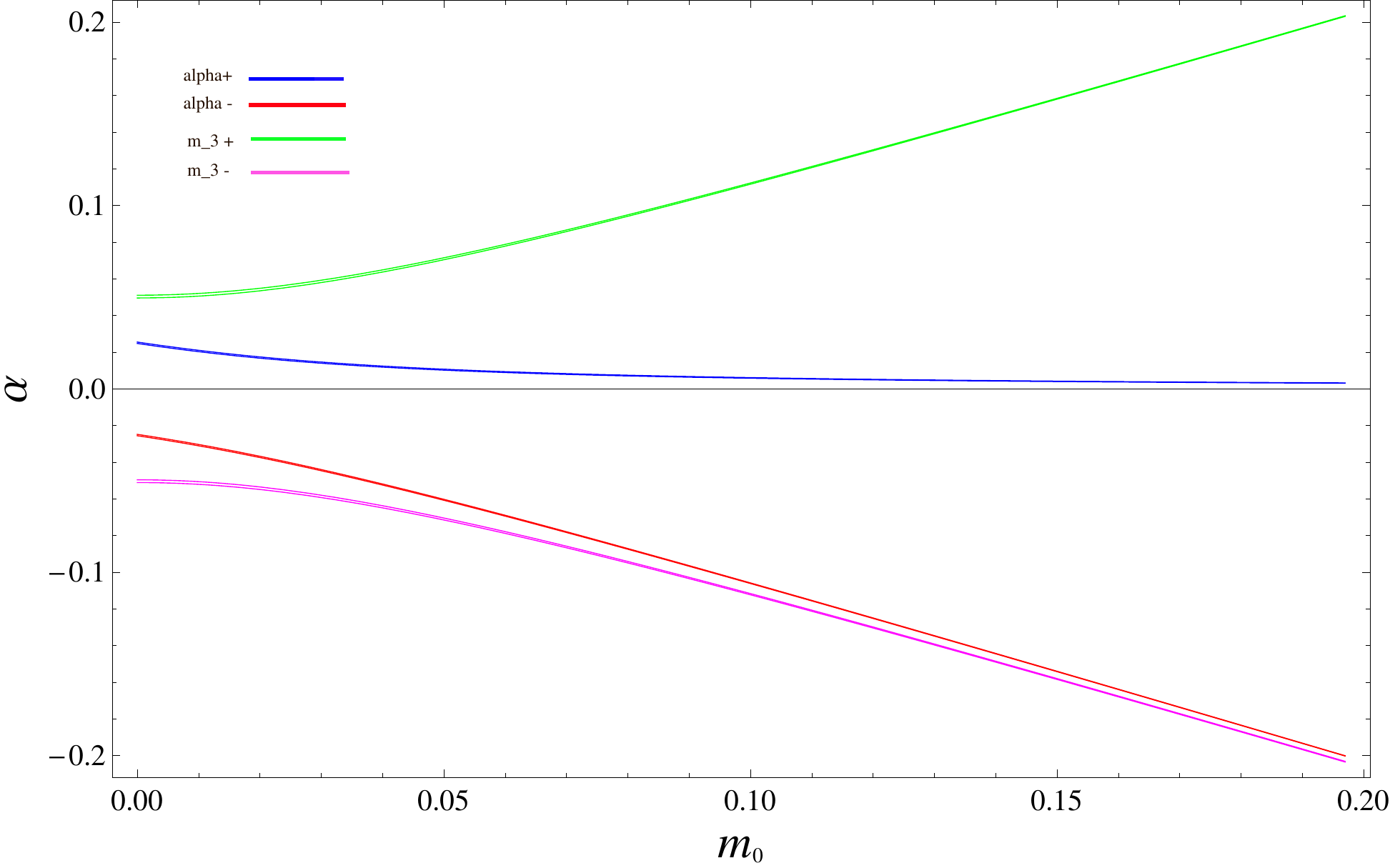}\caption{\label{fig.1}
\small{Allowed range of $\alpha$ in $(\alpha, m_0)$ parameter
space. Two symmetric spaces are associated to $m_3>0$ and $m_3<0$
which yield two different results for $\alpha$.}}
\end{figure}

In the first step, we obtain the allowed ranges for the parameters
of the neutrino mass matrix along with the perturbation term. We
do this by mapping neutrino mass constraints obtained from the
experimental data for $\Delta m_{31}^{2}$ and $\Delta m_{21}^{2}$
onto our parameter space. In figure \ref{fig.1}, we have
 shown the limits imposed by $\Delta m_{31}^{2}$
on the $\alpha$ and $m_0$ parameter space space\footnote{Notice
that the parameters space are the products of the Yukawa coupling
constant and the vacuum expectation value of the Higgs boson. We
should note that their ranges are important. However, these
investigations do not fall within the scope of the present work.}
of our model. However, as seen from Eq.\,(\ref{emm2}), $\Delta
m_{31}^{2}=(2\alpha+m_0)^2-m_0^2$. The values of $\alpha$, coming
from Eq.\,(\ref{alpha}), are
\begin{equation}\label{alpha}
\alpha_{\pm}=-\frac{1}{2}~m_0\pm\frac{1}{2}\sqrt{m_0^2+|\Delta
m_{31}^{2}|},
\end{equation}
  where we have denoted these two solutions with upper (plus) and lower (minus) signs.
  As we see, we always have $\alpha_+>0$ and
  $\alpha_-<0$. In order to visualize the results of Eq.\,(\ref{alpha}), in figure \ref{fig.1}
  we have plotted $\alpha$ versus $m_0$. We see that we can have $m_3 > 0$ or $m_3 < 0$, corresponding to $\alpha_+$ and $\alpha_-$,
  respectively.
     The range of $m_0$ is very impressible and important because when $m_0>0.197$
  then we get $\alpha_+\rightarrow 0$ which yields $m_3^+\rightarrow m_0$ and $\alpha_-\rightarrow -m_0$ which yields $m_3^-\rightarrow -m_0$.
  Such that for both of the cases, the results are unacceptable by comparing with the experimental data. The interesting point is that
  these results show that the physical mass spectrum is identical
for both cases, the green and magenta curves, corresponding to
$m_3^+$ and $m_3^-$ respectively. These curves are symmetric about
the $m_0$ axis which is implying phase choice for the $m_3$s, as
it is seen in Table \ref{h3}. Namely, the Majorana phases are
different for each case. Hence $|m_3^+|=|m_3^-|$, and the value of
$m_3$s in both case is the same. Naturally, the physics in the
perturbation matrix elements does not depend on the chosen
solution. We can see no reason to prefer either specific solution.
For both values of $\alpha$, our model has normal hierarchy, the
same result as in \cite{my}.

\begin{figure}[th] \includegraphics[width=10cm]{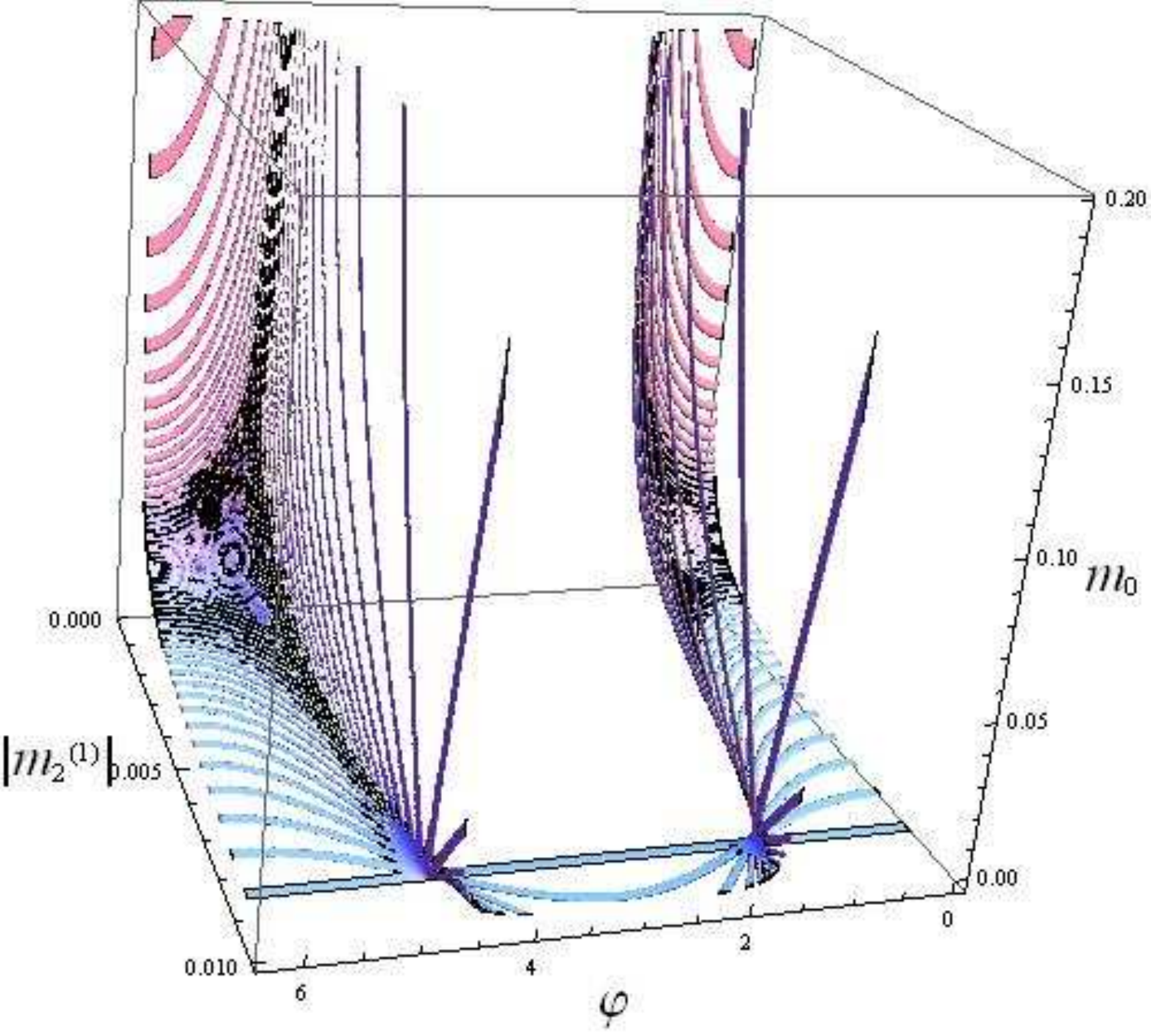}\caption{\label{fig.2}
\small{In this figure, the whole region of the
$|m^{(1)}_2|$-$\varphi$
    plane which is allowed by our model along with the allowed range of $m_0$ is
   shown. Each color curve implies a value of $m_0$ in the rang
    $(0-0.197)eV$in the $|m_2^{(1)}|^2+2m_0m_2^{(1)}\cos\varphi$
    in our model.
  The overlap region of the experimental values for $\Delta
m_{21}^{2}$ with our model are two tiny regions. These regions are
the semi-symmetry of each other.}}
\end{figure}

\begin{figure}[th] \includegraphics[width=13cm]{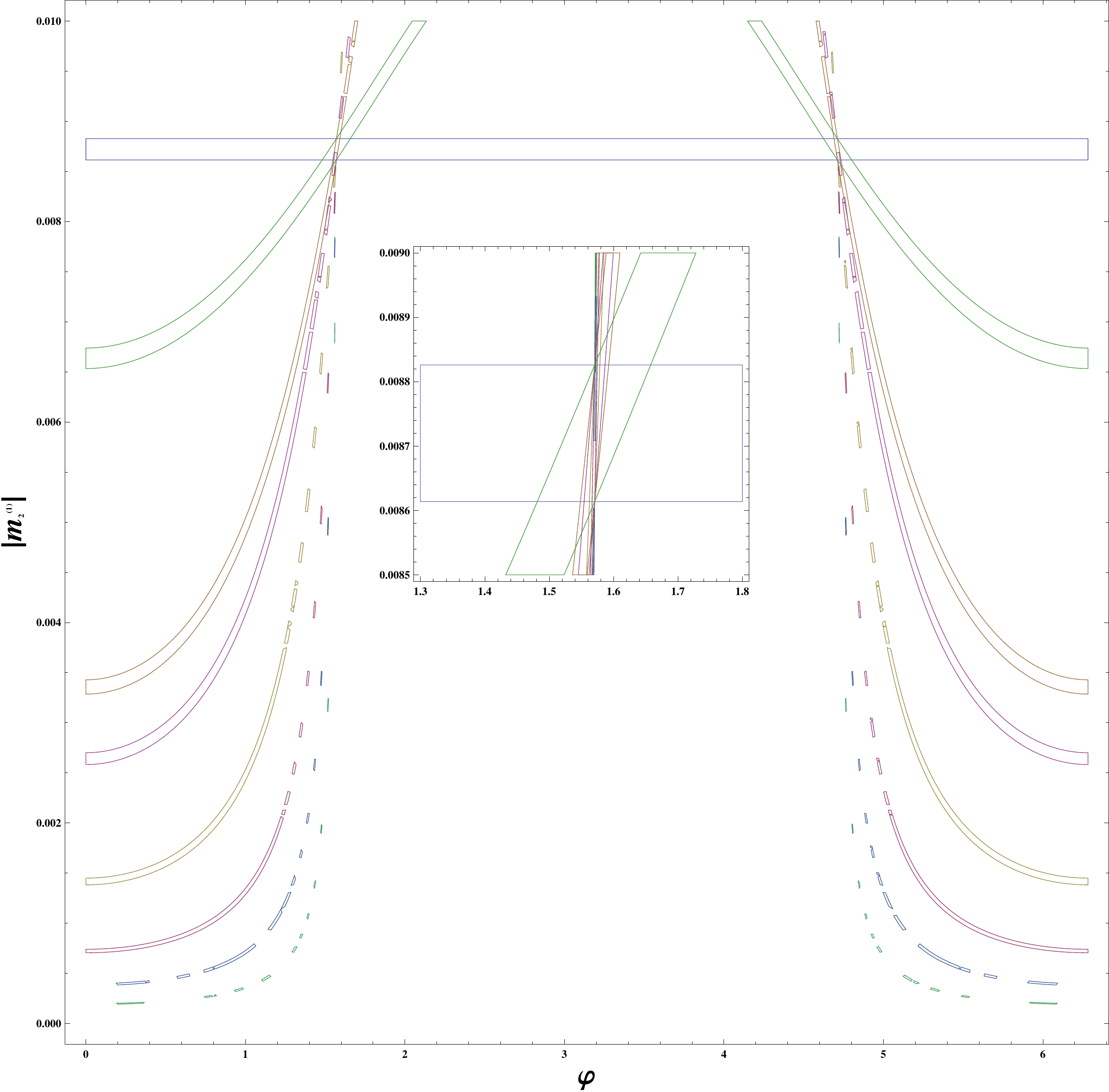}\caption{\label{fig.3}
\small{In this figure, the contour plot of Figure 2, the entire
region of the $|m^{(1)}_2|$-$\varphi$
    plane with the allowed range of $m_0$ are shown. In the zoomed box we have
magnified the right overlap region.}}
\end{figure}

In figure \ref{fig.2} we have plotted the overlap of $\Delta
m_{21}^{2}$, by using Eqs. (\ref{emm2}) and (\ref{em16}), $\Delta
m_{21}^{2}= |m_2^{(1)}|^2+2m_0m_2^{(1)}\cos\varphi$
 with the results obtained in our model along with the allowed ranges of $m_0$ onto
 the $|m^{(1)}_2|$ and $\varphi$ perturbation parameter space in comparison with experimental data.
 In figure \ref{fig.2}, each colored curve implies a value of $m_0$ in the equation $\Delta m_{21}^{2}$ in our model.
  All these curves overlap with each other in a narrow area in the plane of $|m^{(1)}_2|$ and $\varphi$.  Therefore,
 in figure \ref{fig.3}, we have depicted
  the contour plot of figure \ref{fig.2} such that we could clearly show the boundaries of $|m^{(1)}_2|$ and $\varphi$. Our results for
the mass matrix parameters are given by
\begin{eqnarray}\label{eparameter1}
m_{0}&\approx&(0-0.197)eV ,\nonumber\\
\alpha_-&\approx&-(0.0245-0.2)eV,\nonumber\\
\alpha_+&\approx&(0.0245-0.0033)eV,\nonumber\\
|m^{(1)}_2|&\approx&(0.00862-0.00883)eV,\nonumber\\
\varphi&\approx&(89.98^\circ-90.10^\circ).
\end{eqnarray}

Note that the value of $\varphi$ in Eq.\,(\ref{eparameter1}) shows
that $m^{(1)}_2$, the solar neutrino mass splitting term, is
dominated by its imaginary part. Therefore, due to the allowed
range of $\varphi$ [according to (\ref{eparameter1})], the origin
of the Majorana phases, $\phi$ [in Eq.~(\ref{em16})], can take
large values. This seems to suggest that the Majorana nature of
neutrinos can be responsible for a large value of CP violation in
nature~\cite{Majoranaph}.

We expect that the different nonzero components of the
perturbation matrix Eq.\,(\ref{eM2}) are roughly of similar order.
We may then expect
$\varepsilon\equiv\frac{m_2^{(1)}g(\eta)}{|\Delta
m_{31}^{2}|s_{13}}\sim\mathcal{O}(1)$, and could predict the order
of $\sin\theta_{13}$ in $\varepsilon$. Therefore, by using the
order of $\alpha$ and $|m^{(1)}_2|$ from the previous stage, we
obtain $\sin\theta_{13}\sim \mathcal{O}(10^{-1})$.

In the second step, we obtain the allowed ranges for $\delta$ and
$J$, the Jarlskog parameter, as in Eq.\,(\ref{eJ}) and for this we
must first determine the allowed range of $\eta$. For this we use
the expression of $g(\eta)$ in Eq.\,(\ref{ephi}). Recalling
$|m_1|-|m_3|< g(\eta)< |m_1|+|m_3|$, and so $0.049< g(\eta)<
0.400$, we obtain the allowed range of $\eta$ as
$|\eta|\lesssim92.29^\circ$. In the limit $m_0\rightarrow 0$, we
get $\eta\rightarrow \delta$ and $g(\eta)\rightarrow |m_3|$. In
order to get the allowed values of $\delta$, we substitute the
allowed ranges of $m_0$, $\alpha$ and $\eta$ into the expression
associated to $\delta$ in Eq.\,(\ref{edelta}). The results are
\begin{eqnarray}\label{eparameter2}
|\delta|&\lesssim&(21.48^\circ-92.29^\circ) ,\nonumber\\
|J|&\lesssim&(0.012-0.035).
\end{eqnarray}

Not only we have obtained all the parameters of the model
[according to~(\ref{eparameter1})], but now we can also make
predictions for the masses of the neutrinos as well as the phases,
see equation~(\ref{eparameter3}).
 We emphasize that we have made
 predictions that correspond to physical quantities for which there
is yet no experimental data. These predictions include
\begin{eqnarray}\label{eparameter3}
m_{1}&\approx&(0-0.197)eV ,\nonumber\\
|m_2|&\approx&(0.00862-0.19719) eV~~~~~~\text{and}~~~~~~~\phi\approx(2.6^\circ-89.98^\circ),\nonumber\\
|m_3|&\approx&(0.0490-0.2033)eV.
,\nonumber\\
\Delta
m_{21}^{2}&\approx&(7.43-7.49)\times10^{-5}eV,\nonumber\\
\Delta m_{31}^{2}&\approx&(2.40-2.52)\times10^{-3}eV.
\end{eqnarray}
As mentioned, $\phi$ is the origin of the Majorana phases which is
retrieved from the perturbation. We could dispense with the
overall phase, $\exp(i\phi)$, and therefore we would have two
phases that appear in the mass eigenvalues shown in
Eq.\,(\ref{eparameter3}). As we noted, these phases are not the
same for $|m_3|\equiv m_{3\mp}$. Namely, for $m_{3-}$ we have
$\rho=(-\frac{\phi}{2})$ and
$\sigma=(\frac{\pi}{2}-\frac{\phi}{2})$. Whereas, for $m_{3+}$,
$\rho=\sigma=(-\frac{\phi}{2})$. For the Dirac neutrinos, these
phases can be removed, and for the Majorana neutrinos, these
phases remain as Majorana phases and contribute to CP violation.

 As mentioned, in the Appendix A, in the FL setting
$m_0\neq0$, while the magnitude of the lower bound in
Eq.~(\ref{eparameter3}) is about $m_0=0$. These predictions are
compatible with the neutrino mass predictions in \cite{my}.
However, it is important to note that the results in this work
(such as $\Delta m_{21}^{2}$ and $\Delta m_{31}^{2}$), established
by means of a different approach, fit the experimental data much
better than those reported in \cite{my} especially the best fit
column in Table~\ref{h1}. In order to compare our methodology and
approach to that in \cite{my} and employing the current
experimental data, the ratio $\frac{\Delta m_{21}^{2}}{\Delta
m_{31}^{2}}$ can be considered as an appropriate parameter. For
instance, herein we obtain $\frac{\Delta m_{21}^{2}}{\Delta
m_{31}^{2}}\approx(2.97-3.09)\times10^{-2}$, which is closer and
more consistent with the best fit experimental data, $\frac{\Delta
m_{21}^{2}}{\Delta m_{31}^{2}}\approx(2.99-3.02)\times10^{-2}$,
than the corresponding result reported in \cite{my}. Namely, the
agreement rate using our improved framework regarding the best fit
experimental data is about $30$ percent better than in \cite{my}.
Moreover, in \cite{FL}, the phenomenological calculations
associated to the CP violation case have not been compared with
the corresponding experimental data. In Table II, we have
displayed all of the relevant experimental data presented at Table
I along with the predictions of our model. As is shown in Table
II, we have predictions for some physical quantities for which
there are not any experimental data.

\begin{table*}[t]
\begin{center}
\begin{tabular}{|c|c|c|c|}
\hline Parameter & The exp. data    &The best exp.
fit($\pm1\sigma$) &Predictions of our model
\\
\hline
$\Delta m_{21}^{2}(10^{-5}eV^{2})$  & $(6.94-8.14)$ &$(7.30-7.72)$&$(7.43-7.49)$  \\
\hline
$\Delta m_{31}^{2}(10^{-3}eV^{2})$ & $(2.47-2.63)$ &$(2.52-2.57)$&$ (2.40-2.52)$\\
\hline
$\delta$ & ...&... & $\lesssim(21.48^\circ-92.29^\circ)$ \\
\hline
$|J|$ & ...&...&$\lesssim(0.012-0.035)$ \\
\hline
~ &...&...&$m_1\approx(0-0.197)eV,$ \\
~&~&~&$\langle m_{\nu_e} \rangle\approx(0.00386-0.20162)~eV$\\
\cline{2-4}
masses &...&...&$|m_2|\approx(0.00862-0.19719) eV,~\phi\approx(2.6^\circ-89.98^\circ)$ \\
~&~&~&$\langle m_{\nu_\mu} \rangle\approx(0.02737-0.20243)~ eV$\\
\cline{2-4}
~ &...&...&$m_{3\mp}\equiv~|m_3| \approx(0.0490-0.2033)eV,$ \\
~&~&~&$\langle m_{\nu_\tau} \rangle\approx(0.02743-0.20248)~ eV$\\
\hline
$\rho$ and $\sigma$ &...&...&$m_3<0$, $\rho\lesssim-(1.3^\circ-44.99^\circ)$, $\sigma\lesssim(88.70^\circ-45.01^\circ)$ \\
~&~&~&$m_3>0 $,
$\rho=\sigma\lesssim-(1.3^\circ-44.99^\circ)$\\
\hline $\langle m_{\nu_{\beta\beta}} \rangle $ & $ < (0.12 -
0.25)~eV $ at 90\% CL &...&$\langle m_{\nu_{\beta\beta}}
\rangle\approx(0.0028-0.12)~eV$ \\
\hline
\end{tabular}
\caption{The available experimental data for neutrinos for the
case of normal mass hierarchy and the predictions of our model.
These predictions are obtained from our parameters as shown in
Eq.\,(\ref{eparameter1}).}\label{h3}
\end{center}
\end{table*}

In this model, the magnitude of degeneracy associated with the
neutrino masses is defined by $\frac{m_{\rm 3}-m_{\rm 1}}{m_{\rm
3}}$. Hence, the limit $m_{\rm 1}\rightarrow0~(m_{\rm 3})$ means
$0\%~(100\%)$ degeneracy among the neutrino masses~\cite{deg}. We
should note that by using the allowed ranges of $m_1$ and $m_3$ in
Eq.\,(\ref{eparameter3}) the magnitude of degeneracy of the
neutrino masses in our model is $\approx(0\%-97\%)$.

For the flavor eigenstates, we can just calculate the expectation
values associated to the masses. Therefore, we can use
\begin{equation}\label{enu}
\langle m_{\nu_{i} }\rangle=\sum_{j=1}^3|U_{ij}|^2|m_j|,
\end{equation}
where $i=e,~\mu,~\tau$. Our predictions for these quantities are
as follows, $\langle m_{\nu_e} \rangle\approx(0.00386-0.20162)~
eV$, $\langle m_{\nu_\mu} \rangle\approx(0.02737-0.20243)~ eV$ and
$\langle m_{\nu_\tau} \rangle\approx(0.02743-0.20248)~ eV$. The
Majorana neutrinos can violate lepton number, for example in
neutrinoless double beta decay $(\beta\beta0\nu)$
\cite{neutrinoless}. Such a process has not yet been observed and
an upper bound has been set for the relevant quantity, {\em i.e.}
$\langle m_{\nu_{\beta\beta} }\rangle$. Results from the first
phase of the KamLAND-Zen experiment sets the following constraint
$\langle m_{\nu_{\beta\beta}} \rangle < (0.061-0.165)~eV $ at 90\%
CL \cite{kamland}. Our prediction (better than those in \cite{my})
for this quantity is $\langle m_{\nu_{\beta\beta}}
\rangle\approx(0.0028-0.12)~eV$ which is consistent with the
result of kamLAND-Zen experiment.

One of the main experimental result is the sum of the three light
neutrino masses which has just been reported by the
\textit{Planck} measurements of the cosmic microwave background
(CMB) at $95\%$ CL \cite{planck} as
\begin{equation}\label{eplank}\vspace{.2cm}
\sum m_\nu<0.12eV \text{(Plank+WMAP+CMB+BAO)}.
\end{equation}

In our model, we obtain $\sum m_\nu\approx(0.058-0.597)eV$, which
is in agreement with (\ref{eplank}).

\section{Discussion and Conclusions}
\label{concl}
In the next stage, we will consider the attendance
of a small contribution,
 which can be obtained by employing the perturbation theory, which generates small parameters
 in the neutrino mixing component, namely, $U_{13}$, ($\theta_{13}$ and $\delta$), $\Delta m^2_{21}$
  and provides minor amendments to $\theta_{12}$ (but not to $\theta_{23}$).
  CP violation will be investigated.

We should emphasize that in our present work, the method for
retrieving the minimal perturbation mass matrix is completely
different from those present in the literature (see, e.g.,
\cite{my} and references therein) and it can be considered as a
more fundamentally based approach. The distinguishing features of
our herein model are: i) solely from using the third perturbed
mass eigenstate and by employing the rules of the perturbation
theory, we constructed the minimal perturbation matrix of the
basic tribimaximal mixing matrix, producing a modified
Friedberg-Lee model. Therefor, it was produced from the rules of
perturbation theory.\footnote{As it is usual, in the most of the
perturbative analysis, together with some assumptions, a
perturbation matrix is added by hand.} ii) the perturbation mass
matrix is simultaneously responsible for the solar neutrino mass
splitting and CP nonconservation in the lepton sector.
Consequently, due to these two initiatives and distinctive
consideration, our modified framework regarding the best fit
experimental data is 30 percent better than in \cite{my}. The
model is based on the tribimaximal mixing matrix in which the
experimental data of mixing angles (except $\theta_{13}$) is well
approximated. Therefore, by employing the Friedberg-Lee neutrino
mass framework, we obtained the tribimaximal structure which led
us to produce a mass matrix constrained by the elements of the TBM
mixing matrix and the experimental data. The mass matrix thus
obtained (unperturbed mass matrix) loses the solar neutrino mass
splitting whilst it remains as a magic and symmetric matrix under
$\mu-\tau$ symmetry. At this level, by employing perturbation
theory, we generate a perturbation matrix which breaks softly both
the $\mu-\tau$ symmetry and the magic feature, and consequently
causes CP violation.

Our investigation proceeded in two stages [of section \ref{Set
up}]: CP conservation and CP violation. In the first stage, we
obtained the elements of the perturbation matrix in a non CP
violation case. In this case, the elements of the perturbation
matrix are real, and therefore $\delta=0$ while
$\theta_{13}\neq0$.

In the second stage, we extended our study to the case of CP
violation, and we obtained the complex elements of the
perturbation matrix in both the flavor and mass bases. Moreover,
(a) We retrieved a realistic mixing matrix with $\delta\neq0$.
However, in this case, the $\mu-\tau$ symmetry is softly broken,
but still we have $\theta_{23}=45^\circ$. (b) We obtained the
solar neutrino mass splitting dominated by an imaginary term.
Therefore, the most important corresponding results or claims are:
the possibility that Majorana neutrinos exist and are more
abundant than Dirac neutrinos, and, that this feature would be the
reason for the magnitude of CP violation verified in nature.

In order to get valuable predictions concerning neutrino masses,
$\delta$, the origin of the Majorana phases and $J$, we compared
the results of our phenomenological model with experimental data.
We have shown that how our phenomenological model whether or not
is consistent with experimental data. Mapping two sets of
experimental data, namely, the allowed ranges of $\Delta m^2_{21}$
and $\Delta m^2_{31}$ onto the allowed region of our parameter
space can determine valid values for our parameters. The
consistency of experimental data with the allowed ranges of our
parameters shows that our model has normal hierarchy. We then
predict the perturbation mass matrix and the values of three
masses, $m_{1}\approx(0-0.197)eV $, $|m_2|\approx(0.00862-0.19719)
eV$, and $m_3\approx\mp(0.0490-0.2033)eV$. Therefore, the
magnitude of degeneracy for neutrino masses at the end of the
allowed ranges is about $97\%$. We have shown that the order of
$\sin\theta_{13}$, which is estimated from the order of the mass
parameters, is consistent with the experimental data. We also
obtain predictions for the CP violation parameters $\delta$, $J$,
$\rho$ and $\sigma$. These are
$|\delta|\lesssim(21.48^\circ-92.29^\circ)$,
$|J|\lesssim(0.012-0.035)$, while the values of the Majorana
phases depend on the sign of $m_3$: for $m_3<0$,
$\rho\lesssim-(1.3^\circ-44.99^\circ)$,
$\sigma\lesssim(88.70^\circ-45.01^\circ)$, while for $m_3>0 $,
$\rho=\sigma\lesssim-(1.3^\circ-44.99^\circ)$. Our predictions for
the neutrino masses and CP violation parameters could be tested in
future experiments such as the upcoming long baseline neutrino
oscillation ones. Moreover, our predictions are entirely
consistent with the constraints reported by Planck, WP and high L
measurements and the KamLAND-Zen experiment
\cite{planck,neutrinoless}.

In our model, the minimal perturbation matrix was obtained by
means of a fundamental process, and not merely added by hand.
Therefore, we can claim that the model presented here can be
regarded as presenting a more  comprehensive scenario. Another
important point is that although our predictions are in
correspondence with those
in~\cite{my}, we should emphasize that our outcomes (such as
$\Delta m_{21}^{2}$ and $\Delta m_{31}^{2}$), constitute with the
best fit experimental data reported in table~\ref{h1}. Last and
not least to emphasize, the consistency fitting rate concerning
the best fit experimental data is about $30$ percent more
efficient trough the framework introduced in this paper rather
than in \cite{my}.

 Finally, we plan to subsequently proceed and
investigate a $6\times6$ neutrino mass matrix, by using the same
methodology, to obtain the corresponding perturbation mass matrix
and CP violation.

\section{Acknowledgments}
acknowledge the FCT grants UID-B-MAT/00212/2020 and
UID-P-MAT/00212/2020 at CMA-UBI.

\appendix
\section{The Fridberg-Lee model}
\label{App.A}

In the FL model, the mass eigenstates of the three charged leptons
are the same as their corresponding flavor eigenstates. Therefore,
the neutrino mixing matrix is simply the $3\times3$ unitary matrix
$U$, which transforms the neutrino mass eigenstates to the flavor
eigenstates $ (\nu_{e},\nu_{\mu},\nu_{\tau})$. The neutrino mass
operator can be written as \cite{FL}
\begin{eqnarray}\label{emfl}\vspace{.5cm}
  {{\cal M}_{FL}}&=&a\left(\bar{\nu}_{\tau}-\bar{\nu}_{\mu}\right)\left(\nu_{\tau}-\nu_{\mu}\right)
+
b\left(\bar{\nu}_{\mu}-\bar{\nu}_{e}\right)\left(\nu_{\mu}-\nu_{e}\right)
+ c\left(\bar{\nu}_{e}-\bar{\nu}_{\tau}\right)\left(\nu_{e}-\nu_{\tau}\right)\nonumber\\
&+&
m_{0}\left(\bar{\nu}_{e}\nu_{e}+\bar{\nu}_{\mu}\nu_{\mu}+\bar{\nu}_{\tau}\nu_{\tau}\right).
\end{eqnarray}
 All the parameters ($a,b,c$ and
$m_{0}$) are assumed to be real. In the original FL setup, also
known as the pure FL model, $m_0=0$ and consequently ${\cal
M}_{FL}$ admits the following symmetry
$\nu_{e}\rightarrow\nu_{e}+z $, $\nu_{\mu}\rightarrow\nu_{\mu}+z
$, and $ \nu_{\tau}\rightarrow\nu_{\tau}+z $~\cite{FL}, where $z$
is an element of the Grassman algebra. When $z$ is a constant,
this is called the FL symmetry~\cite{FL}, and the kinetic term is
also invariant, but the other terms of the electroweak Lagrangian
do not exhibit this symmetry. The $ m_{0} $ term  breaks this
symmetry explicitly. However, we may add that the FL symmetry
leads to a magic matrix~\cite{magic} and this property is not
spoiled by the $m_{0}$ term~\cite{FL}. It has also been argued
that the FL symmetry is the residual symmetry of the neutrino mass
matrix after the $SO(3)\times U(1)$ flavor symmetry breaking
\cite{FL2}. The mass matrix can be displayed as~\cite{FL}
\begin{equation}\label{efl}
M_{FL} =\left(\begin{array}{ccc}b+c+m_{0} & -b & -c\\
-b & a+b+m_{0} & -a\\-c & -a & a+c+m_{0}\end{array}\right),
\end{equation}
where $a \propto\left(Y_{\mu\tau}+Y_{\tau\mu }\right)$, $b
\propto\left(Y_{e\mu}+Y_{\mu e }\right)$ and $  \ c
\propto\left(Y_{\tau e}+Y_{e\tau }\right) $ and $Y_{\alpha\beta}$
denote the Yukawa coupling constants~\cite{FL}. Notice that $
M_{FL}$ in Eq.\,(\ref{efl}) is symmetric, and therefore could be
used for Dirac or for Majorana neutrino mass terms. The
proportionality constant is the expectation value of the Higgs
field. As mentioned, it is clear that the first three terms in
Eq.\,(\ref{emfl}) are invariant under the transformation
$\nu_{\alpha}\rightarrow\nu_{\alpha}+z $ (for
$\alpha=e,~\mu,~\tau$). The same invariance can also be expressed
in terms of the transformation between the constants, $a$, $b$,
and $c$, with
\begin{equation}\label{elanda}
a\rightarrow a+\lambda,~~~~~~b\rightarrow
b+\lambda,~~~~~~\text{and}~~~~~~c\rightarrow c+\lambda.
\end{equation}
Therefore, under the transformations~(\ref{elanda}), the form of
the neutrino mixing matrix $U$ remains unchanged~\cite{FL}.

In order to have CP-violation, within the standard parametrization
given by Eq.\,(\ref{emixing}), the necessary condition is
$\delta\neq0$ and $\theta_{13}\neq0$. There are four independent
CP-even quadratic invariants, which can conveniently be chosen as
$U^{\ast}_{11}U_{11}, U^{\ast}_{13}U_{13}, U^{\ast}_{21}U_{21}$
and $ U^{\ast}_{23}U_{23} $ and three independent CP-odd quartic
invariants~\cite{quadratic},
\begin{eqnarray}\label{eJ}\vspace{.2cm}
J&=&{\cal{I}}m(U_{11}U^{\ast}_{12}U^{\ast}_{21}U_{22})\nonumber\\
I_{1}&=&{\cal{I}}m[(U^{\ast}_{11}U_{12})^{2}]\nonumber\\I_{2}&=&{\cal{I}}m[(U^{\ast}_{11}U_{13})^{2}].
\end{eqnarray}
The Jarlskog rephasing invariant parameter $J$~\cite{J} is
relevant to CP violation in lepton number conserving processes
like neutrino oscillations. $I_{1}$ and $I_{2}$ are relevant to CP
violation in lepton number violating processes like neutrinoless
double beta decay. Oscillation experiments cannot distinguish the
Dirac from the Majorana neutrinos~\cite{dm}. The detection of
neutrinoless double beta decay would provide direct evidence of
lepton number non-conservation and the Majorana nature of
neutrinos. Many theoretical and phenomenological investigations
have discussed neutrino mass models which break $\mu-\tau$
symmetry as a prelude to CP violation~\cite{theoretical1,
theoretical2, theoretical3}.


\begin{thebibliography}{99}

\bibitem{my}
N. Razzaghi, S. S. Gousheh, Phys. Rev. D 89, 033010 (2014).
\bibitem{exp1}
SNO Collaboration, Q.R. Ahmad, et al., Direct Evidence for
Neutrino Flavor Transformation from Neutral-Current Interactions
in the Sudbury Neutrino Observatory, Phys. Rev. Lett. 89 (2002)
011301.
\bibitem{exp2}
KamLAND Collaboration, K. Eguchi, et al., First Results from
KamLAND: Evidence for Reactor Antineutrino Disappearance, Phys.
Rev. Lett. 90 (2003) 021802.
\bibitem{exp3}
 K2K Collaboration, M.H. Ahn, et al., Indications of Neutrino Oscillation in a
 250 km Long-Baseline Experiment, Phys. Rev.
Lett. 90 (2003) 041801.
\bibitem{exp4}
D. A. Dwyer [Daya Bay Collaboration], The Improved Measurement of
Electron-antineutrino Disappearance at Daya Bay, Nucl. Phys. Proc.
Suppl. 235-236, 30 (2013) [hep-ex/1303.3863].
\bibitem{information}
 P. F. de Salas, D. V. Forero, S. Gariazzo, P. Martinez-Mirave, O.
Mena, C. A. Ternes, M. Tortola, J. W. F. Valle, J. High Energ.
Phys. 2021, 71 (2021).
\bibitem{mixing1}
J. Schechter and J. W. F. Valle, Neutrino masses in
$SU(2)\bigotimes U(1)$ theories, Phys.Rev. D22 (1980) 2227.
\bibitem{mixing2}
 H. Fritzsch and Z.Z. Xing, How to Describe Neutrino Mixing and CP Violation
 , Phys. Lett. B 517, 363 (2001) [hep-ph/0103242v2].
\bibitem{mixing3}
Particle Data Group, W.M. Yao et al., Review of Particle Physics ,
J. Phys. G 33, 1 (2006).
\bibitem{TBM1}
P. F. Harrison at al., A Redetermination of the Neutrino
Mass-Squared Difference in Tri-Maximal Mixing with Terrestrial
Matter Effects, Phys. Lett. B458, 79 (1999) [hep-ph/9904297v1].
\bibitem{TBM2}
P. F. Harrison et al., Tri-Bimaximal Mixing and the Neutrino
Oscillation Data, Phys. Lett. B530, 167 (2002) [hep-ph/0202074v1]; Z.~z.~Xing,
``Nearly tri bimaximal neutrino mixing and CP violation,''
Phys. Lett. B \textbf{533}, 85-93 (2002)
[arXiv:hep-ph/0204049 [hep-ph]].
\bibitem{TBM3}
 X.-G. He and A. Zee, Some Simple Mixing and Mass Matrices for Neutrinos, Phys.
Lett. B560, 87 (2003) [hep-ph/0301092v3].
\bibitem{81}
P. Ciafaloni, M. Picariello, A. Urbano and E. Torrente-Lujan,
Toward a minimal renormalizable supersymmetric SU(5) grand unified
model with tribimaximal mixing from $A_4$ flavor symmetry, Phys.
Rev. D 81 ,016004 (2010) [hep-ph/0909.2553].
\bibitem{82}
 P. H. Frampton , T. W. Kephart and S. Mat- suzaki,
 Simplified renormalizable T' model for tribimaximal mixing and Cabibbo
 angle, Phys. Rev. D 78 , 073004 (2008) [hep-ph/0807.4713].
 \bibitem{83}
 F. Plentinger , G. Seidl and W. Winter, Group space scan of flavor symmetries for
 nearly tribimaximal lepton mixing, JHEP 0804 , 077 (2008)
[hep-ph/0802.1718].
 \bibitem{84}
 F. Bazzocchi, S. M orisi and M.
Picariello, Embedding A4 into left-right flavor symmetry:
Tribimaximal neutrino mixing and fermion hierarchy, Phys. Lett. B
659 , 628 (2008) [hep-ph/0710.2928].
\bibitem{85}
G. Altarelli and F. Feruglio, Tri-Bimaximal Neutrino Mixing, A4
and the Modular Symmetry, Nucl. Phys. B 741 , 215 (2006)
[hep-ph/0512103].
\bibitem{855}
Z.~z.~Xing,
J. Phys. G \textbf{49}, no.2, 025003 (2022)
[arXiv:2102.03050 [hep-ph]].
\bibitem{HeZee}
X. He, A. Zee, Minimal modification to tribimaximal mixing, Phys.
Rev. D84, 053004 (2011).
\bibitem{final}
Biswajoy Brahmachari, Amitava Raychaudhuri, Perturbative
generation of $\theta_{13}$ from tribimaximal neutrino mixing,
Phys. Rev. D 86, 051302(R) (2012) [hep-ph/1204.5619v3].
\bibitem{Grinstein}
B.~Grinstein and M.~Trott, [hep-ph/1203.4410].
\bibitem{unitary1}
  S.~F.~King, Parametrizing the lepton mixing matrix in terms of deviations from tri-bimaximal
  mixing, Phys.\ Lett.\ B {\bf 659}, 244 (2008) [hep-ph/0710.0530].
\bibitem{unitary2}
S. Pakvasa,   W. Rodejohann, T. Weiler, Unitary Parametrization of
Perturbations to Tribimaximal Neutrino Mixing, Phys. Rev. Lett.
{\bf 100}, 111801 (2008).
\bibitem{unitary3}
C. H. Albright, A. Dueck, W. Rodejohann, Possible Alternatives to
Tri-bimaximal Mixing, Eur. Phys. J.\ C {\bf 70}, 1099-1110 (2010)
[hep-ph/1004.2798v1].
\bibitem{Boudjemaa1}
  S.~Boudjemaa and S.~F.~King, Deviations from Tri-bimaximal Mixing: Charged Lepton Corrections and
  Renormalization Group Running, Phys.Rev. D 79, 033001 (2009) [hep-ph/0808.2782].
\bibitem{Boudjemaa2}
  S.~Goswami, S.~T.~Petcov, S.~Ray and W.~Rodejohann, Large $U_{e3}$ and Tri-bimaximal
  Mixing,Phys.Rev.D80,053013(2009).
\bibitem{Boudjemaa3}
  D.~Meloni, F.~Plentinger and W.~Winter, Perturbing exactly tri-bimaximal neutrino mixings with charged lepton mass matrices,
Phys. Lett. B  699, 354 (2011) [hep-ph/1012.1618].
\bibitem{Boudjemaa33}
Sumit K. Garg, Consistency of perturbed Tribimaximal, Bimaximal
and Democratic mixing with Neutrino mixing data ,  Nucl. Phys.
B931C (2018) 469-505.
\bibitem{Boudjemaa4}
  D.~Marzocca, S.~T.~Petcov, A.~Romanino and M.~Spinrath, Sizeable $\theta_13$ from the Charged Lepton Sector in SU(5),
  (Tri-)Bimaximal Neutrino Mixing and Dirac CP Violation,
JHEP {\bf 1111}, 009 (2011) [hep-ph/1108.0614].
\bibitem{A4pert1}
  G.~Altarelli and F.~Feruglio, Tri-Bimaximal Neutrino Mixing, A4 and the Modular Symmetry,
Nucl.\ Phys.\ B {\bf 741}, 215 (2006) [hep-ph/0512103].
\bibitem{A4pert2}
  F.~Bazzocchi, S.~Morisi and M.~Picariello, Embedding A4 into left-right flavor symmetry: Tribimaximal neutrino mixing and fermion hierarchy,
Phys.\ Lett.\ B {\bf 659}, 628 (2008) [hep-ph/0710.2928].
\bibitem{A4pert3}
  E.~Ma and D.~Wegman, Nonzero theta(13) for neutrino mixing in the context of A(4) symmetry,
Phys.\ Rev.\ Lett.\  {\bf 107}, 061803 (2011) [hep-ph/1106.4269].
\bibitem{A4pert4}
  S.~Gupta, A.~S.~Joshipura and K.~M.~Patel, Minimal extension of tri-bimaximal mixing and generalized $Z_2$ X $Z_2$ symmetries,
Phys.\ Rev.\ D {\bf 85}, 031903 (2012) [hep-ph/1112.6113].
\bibitem{Adhikary}
  B.~Adhikary, B.~Brahmachari, A.~Ghosal, E.~Ma and M.~K.~Parida,
  $A_4$ symmetry and prediction of $U_{e3}$ in a modified Altarelli-Feruglio model,
Phys.\ Lett.\ B {\bf 638}, 345 (2006) [hep-ph/0603059].
\bibitem{Delta}
  E.~Ma, Near Tribimaximal Neutrino Mixing with $\Delta(27)$ Symmetry, Phys.\ Lett.\ B {\bf 660}, 505 (2008) [hep-ph/0709.0507].
\bibitem{Group1}
  F.~Plentinger, G.~Seidl and W.~Winter, Group space scan of flavor symmetries for nearly tribimaximal lepton mixing,
JHEP {\bf 0804}, 077 (2008) [hep-ph/0802.1718].
\bibitem{Group2}
  N.~Haba, R.~Takahashi, M.~Tanimoto and K.~Yoshioka, Tri-bimaximal Mixing from Cascades,
Phys.\ Rev.\ D {\bf 78}, 113002 (2008) [hep-ph/0804.4055].
\bibitem{Group3}
  S.~-F.~Ge, D.~A.~Dicus and W.~W.~Repko, Residual Symmetries for Neutrino Mixing with a Large $\text{theta}_{13}$ and Nearly Maximal $\text{delta}_D$,
Phys.\ Rev.\ Lett.\  {\bf 108}, 041801 (2012) [hep-ph/1108.0964].
\bibitem{Group4}
  T.~Araki and Y.~F.~Li, ${\text{Q}}_6$ flavor symmetry model for the extension of the minimal standard model by three right-handed sterile neutrinos,
Phys.\ Rev.\ D {\bf 85}, 065016 (2012) [hep-ph/1112.5819].
\bibitem{P1}
F. Vissani, J. High Energy Phys. 9811 (1998) 025; E.K.Akhmedov,
Phys. Lett. B 467 (1999) 95; M. Lindner, W. Rodejohann, J. High
Energy Phys. 0705 (2007) 089; D. Aristizabal Sierra, I. de
Medeiros Varzielas, E. Houet, Phys. Rev. D 87 (2013) 093009; T.
Araki, Prog. Theor. Exp. Phys. 2013 (2013) 103B02; M.-C. Chen, J.
Huang, K.T. Mahanthappa, A.M. Wijangco, J. High Energy Phys. 1310
(2013) 112; L.J. Hall, G.G. Ross, J. High Energy Phys. 1311 (2013)
091; Jiajun Liao, D. Marfatia, K. Whisnant, Phys. Rev. D 92,
073004 (2015) ; Sumit K. Garg,  Nucl. Phys. B931C (2018) 469-505;  Sumit K. Garg, 
Int.J.Mod.Phys.A 36 (2021) 18, 2150118. 
\bibitem{FL}
R. Friedberg, T.D. Lee, A Possible Relation between the Neutrino
Mass Matrix and the Neutrino Mapping Matrix, High Energy Phys.
Nucl. Phys. 30 (2006) 591, arXiv:[hep-ph/0606071].
\bibitem{mutau}
T. Fukuyama and H. Nishiura, arXiv:hep-ph/9702253; R. N. Mohapatra
and S. Nussinov, Phys. Rev. D 60, 013002 (1999); K. R. S. Balaji,
W. Grimus, and T. Schwetz, Phys. Lett. B 508, 301 (2001); C. S.
Lam, Phys. Lett. B 507, 214 (2001);W. Grimus and L. Lavoura, J.
High Energy Phys. 07 (2001) 045.
\bibitem{magic}
C.S. Lam, Phys.Lett. B640 (2006) 260-262[ arXiv:hep-ph/0606220v2].
\bibitem{TTBM}
R. R. Gautam, S. Kumar, Zeros in the magic neutrino mass matrix,
Phys. Rev. D 94, 036004 (2016), arXiv:1607.08328v2 [hep-ph].
\bibitem{FL2}
C. S. Huang, T. Li, W. Liao and S. H. Zhu, Generalization of
Friedberg-Lee symmetry, Phys. Rev. D 78, 013005 (2008).
\bibitem{quadratic}
Elizabeth Jenkins and Aneesh V. Manohar, Rephasing Invariants of
Quark and Lepton Mixing Matrices, Nucl. Phys. B792, 187 (2008).
\bibitem{J}
C. Jarlskog, Commutator of the Quark Mass Matrices in the Standard
Electroweak Model and a Measure of Maximal CP Nonconservation,
Phys. Rev. Lett. 55 (1985) 1039.
\bibitem{dm}
M. Zralek, Acta Phys. Pol. B 41, 2563 (2011).
\bibitem{theoretical1}
Z. Z. Xing, H. Zhang, S. Zhou, Nearly Tri-bimaximal Neutrino
Mixing and CP Violation from mu-tau Symmetry Breaking, Phys Lett B
641 (2006) 189.
\bibitem{theoretical2}
 T.
Baba, M. Yasue, Correlation between Leptonic CP Violation and
mu-tau Symmetry Breaking, Phys. Rev. D75, 055001 (2007).
\bibitem{theoretical3}
 Z. Z. Xing, H.
Zhang, S. Zhou, Generalized Friedberg-Lee model for neutrino
masses and leptonic CP violation from mu-tau symmetry breaking,
Int. J. Mod. Phys. A 23, 3384(2008).
\bibitem{Schiff} L. I. Schiff, {\em Quantum Mechanics} (Third ed.),
McGraw-Hill (1968).
\bibitem{18}
Z.~-z.~Xing, A Shift from Democratic to Tri-bimaximal Neutrino
Mixing with Relatively Large $\text{ theta}_{13}$, Phys. Lett. B
696, 232 (2011) [hep-ph/1011.2954v2].

\bibitem{Majoranaph}
Peter Ballett, Silvia Pascoli, Jessica Turner, Mixing angle and
phase correlations from A5 with generalised CP and their prospects
for discovery, Phys. Rev. D 92, 093008 (2015), arXiv:1503.07543v1
[hep-ph].
\bibitem{deg}
Naoyuki Haba and RyoTakahashi, Constraints on neutrino mass
ordering and degeneracy from Planck and neutrino-less double beta
decay, Phys.Polon. B45 (2014) 1, 61-69 [hep-ph/1305.0147v3].

\bibitem{neutrinoless}
 L. Wolfenestein, Neutrino oscillations in matter, Phys. Rev. D 17, 2369 (1978).
\bibitem{kamland}
Yoshihito Gando [KamLAND-Zen Collaboration], (2018)
[hep-ex/1904.06655v1].
\bibitem{planck}
N. Aghanim et al. [Planck], Astron. Astrophys. 641, A6 (2020),
[arXiv:1807.06209 [astro-ph.CO]].








\end{thebibliography}
\end{document}